\documentclass[12pt]{iopart}
\usepackage{epsfig}

\def\beq{\begin{equation}}
\def\eeq{\end{equation}}
\def\be{\begin{equation}}
\def\ee{\end{equation}}
\def\beaq{\begin{eqnarray}}
\def\eeaq{\end{eqnarray}}
\def\bea{\begin{eqnarray}}
\def\eea{\end{eqnarray}}

\newcommand{\xbj}{x_{\scriptscriptstyle B}}
\newcommand{\bpt}{\bm p_\sT}

\newcommand{\psibar}{\overline{\psi}}

\newcommand{\slsh}[1]{\mbox{$\not\! #1$}}
\newcommand{\bm}[1]{\mbox{\boldmath${#1}$}}

\newcommand{\st}{{\scriptscriptstyle T}}
\newcommand{\sT}{{\scriptscriptstyle T}}
\newcommand{\sL}{{\scriptscriptstyle L}}

\begin{document}

\title[Azimuthal asymmetries in semi-inclusive leptoproduction]
{Azimuthal asymmetries in semi-inclusive leptoproduction\footnote{
Talk presented at the Ringberg Workshop: New Trends in HERA Physics,
2001, Ringberg Castle, Germany, June 17-22, 2001.}}

\author{P.J. Mulders
\footnote{mulders@nat.vu.nl}
}

\address{Department for Theoretical Physics, Faculty of Sciences,
Vrije Universiteit,\\
De Boelelaan 1081, 1081 HV Amsterdam,
the Netherlands}

\begin{abstract}
We point out the role of intrinsic transverse momentum of partons
in the study of azimuthal asymmetries in deep-inelastic 1-particle 
inclusive leptoproduction. Leading asymmetries often appear in 
combination with spin asymmetries. This leads not only to
transverse momentum dependence in the parton distribution
functions, but also to functions beyond the ones known from inclusive
deep-inelastic scattering (DIS).
We use Lorentz invariance and the QCD equations of motion to study
the evolution of functions that appear at leading (zeroth) order in a $1/Q$
expansion in azimuthal asymmetries. 
This includes the evolution equation of the Collins fragmentation function. 
The moments of these functions are matrix elements of known twist two 
and twist three operators.
We give the evolution in the large $N_c$ limit, restricted to
the non-singlet case for the chiral-even functions.
\end{abstract}



\maketitle

\section{Inclusive deep-inelastic scattering}

In inclusive deep-inelastic scattering (DIS) the neutral current (
photon exchange) cross section can be expressed in
terms of two structure functions $F_1(\xbj, Q^2)$ and $F_2(\xbj,Q^2)$, which
if the momentum transfer squared $Q^2$ is large can be expressed as a
weighted sum over quark distributions $q(x)$ for the various flavors,
$2\,F_1(\xbj, Q^2)$ = $F_2(\xbj, Q^2)/\xbj$ = $\sum_q e_q^2\,q(\xbj)$. In 
polarized deep-inelastic scattering with both lepton and target polarized 
again two structure functions $g_1(\xbj,Q^2)$ and $g_2(\xbj,Q^2)$ appear,
of which for large $Q^2$ the first one can be expressed as a weighted sum over
polarized quark distributions $\Delta q(x)$ for the various flavors.

In hard processes such as DIS the effects of hadrons can be studied via quark and gluon
correlators. In inclusive deep inelastic scattering (DIS), these are 
lightcone\footnote{
For inclusive leptoproduction the lightlike directions $n_\pm$ and lightcone
coordinates $a^\pm = a\cdot n_\mp$ are defined through hadron momentum $P$
and the momentum transfer $q$,
\begin{eqnarray*}
P = \frac{Q}{\xbj\sqrt{2}}\,n_+ + \frac{\xbj M^2}{Q\sqrt{2}}\,n_-,
\\
q = -\frac{Q}{\xbj\sqrt{2}}\,n_+ + \frac{Q}{\sqrt{2}}\,n_-.
\end{eqnarray*}
}
correlators depending on $x \equiv p^+/P^+$ of the type
\beaq
\Phi_{ij} (x) &\equiv& \left. \int \frac{d \xi^-}{2\pi}\ e^{i\,p\cdot
\xi}\langle P,S|\, \psibar_j (0) \,{\cal U}(0,\xi)
\,\psi_i(\xi) | P,S\rangle\right|_{LC} .
\label{PhiDIS}
\eeaq
where the subscript `LC' indicates $\xi^+ = \xi_\sT = 0$ and
${\cal U}(0,\xi)$ is a gauge link with the path running along the minus
direction. The use of lightlike directions enters naturally in a deep-inelastic
process in which the hadrons, compared to the scale $Q^2$, are massless.

The parametrization for DIS at leading (zeroth) order in a
$1/Q$ expansion is 
\beq
\Phi^{{\rm twist}-2}(x)=\frac{1}{2} \left\{{f_1(x)} \slsh n_+ 
+ S_\sL\,{g_1(x)}\,\gamma_{5}\slsh n_+
+ {h_1(x)}\,\gamma_{5}\slsh S_\sT\slsh n_+ \right\} ,
\eeq
where longitudinal spin $S_\sL$ refers to the component along the same
lightlike direction as defined by the hadron\footnote{
The spin vector is parametrized $S = S_\sL\,\frac{P^+}{M}n_+
- S_\sL\,\frac{M}{2P^+}\,n_- + S_\sT$.}. Specifying also the flavor one
also encounters the notations $q(x) = f_1^q(x)$,  $\Delta q(x) = g_1^q(x)$ and
$\delta q(x) = \Delta_T q(x) = h_1^q(x)$. The evolution equations for these
functions are known to next-to-leading order and for the singlet $f_1$ and
$g_1$ there is mixing with the unpolarized and polarized gluon distribution
functions $g(x)$ and $\Delta g(x)$, respectively. 

Instead of using the compact Dirac notation, it is useful to write explicitly
the production matrix $M^{\rm prod}=(\Phi\gamma^+)^T$ that appears in leading 
order DIS calculations.
In hard processes only two Dirac components are relevant, {\em one}
of them righthanded and {\em one} lefthanded ($\psi_{R/L} =
\frac{1}{2}(1\pm \gamma_5)\psi$). The advantage of the matrix representation
is that
$M^{\rm prod}$ is a matrix of which any diagonal element corresponds to a
positive definite momentum distribution. Restricting
ourselves to the two relevant (right-/lefthanded) quark states and using
instead of the spin
vector $S$ explicitly two nucleon spin states, the matrix in the
$4 \times 4$ quark $\otimes$ nucleon spin space
becomes~\cite{BBHM}
\bea
M^{\rm (prod)}\ & = &
\left\lgroup \begin{array}{cccc}
f_1 + g_1 & 0 & 0 & 2\,h_1 \\
& & &\\
0 & f_1 - g_1 & 0 & 0 \\
& & &\\
0 & 0 & f_1 - g_1 & 0 \\
& & &\\
2\,h_1 & 0 & 0 & f_1 + g_1
\end{array}\right\rgroup 
\ \begin{array}{c}
\includegraphics[width = 0.8 cm]{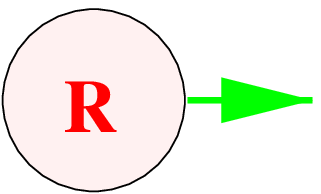}\\[0.3cm]
\includegraphics[width = 0.8 cm]{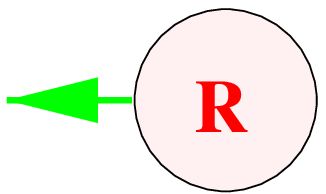}\\[0.3cm]
\includegraphics[width = 0.8 cm]{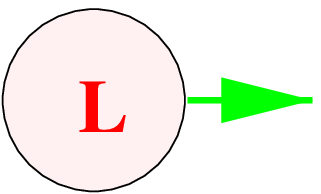}\\[0.3cm]
\includegraphics[width = 0.8 cm]{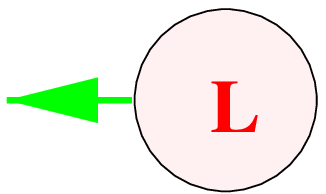}
\end{array}
\label{prod1}
\\
&&\mbox{}\hspace{0.7cm}
\includegraphics[width = 0.8 cm]{mulders.helrr.eps}
\hspace{0.7cm}\includegraphics[width = 0.8 cm]{mulders.hellr.eps}
\hspace{1.0cm}\includegraphics[width = 0.8 cm]{mulders.helrl.eps}
\hspace{0.7cm}\includegraphics[width = 0.8 cm]{mulders.helll.eps}
\nonumber
\eea
From the matrix the physical interpretation of $f_1$ as quark momentum density 
and the interpretation of $g_1$ as the difference of chiral densities (often
sloppily referred to as helicity distribution) is seen. 
As can be seen the function $h_1(x)$ involves a matrix elements between left- 
and right-handed quarks, it is chirally odd~\cite{JJ92}. This implies that it is
not accessible in inclusive DIS, where the hard scattering part does not
change chirality except via (irrelevant) quark mass terms. This third
distribution function, however, is needed for the complete characterization 
of the (collinear) spin state of a proton as probed in hard scattering processes.
By choosing a different basis (actually of transversely polarized quarks and
transversely polarized nucleons) $h_1$ appears on the diagonal and its 
interpretation as tranverse spin polarization in a transversely polarized
target (often referred to as transversity) is seen.

From the fact that any forward matrix element of the above matrix represents a
density, one derives positivity bounds
\bea
&& f_1(x) \ge 0 \\
&& \vert g_1(x)\vert \le f_1(x) \\
&& \vert h_1(x)\vert \le 
\frac{1}{2}\left( f_1(x) + g_1(x)\right) \le f_1(x).
\eea
For the last bound~\cite{Soffer95} one explicitly
needs the semi-definiteness of the matrix.

For DIS up to order $1/Q$ one needs also the 
$M/P^+$ parts in the parameterization of $\Phi(x)$.
Not imposing time-reversal invariance one obtains
\beaq
\Phi^{{\rm twist}-3}(x) &=& \frac{M}{2P^+} \left\{{e(x)} 
+ {g_T(x)}\,\gamma_{5}\slsh S_\sT
+ S_\sL\,{h_L(x)}\,\gamma_{5}\frac{[\slsh n_+,\slsh n_-]}{2} \right\} 
\nonumber \\ 
& + & \frac{M}{2P^+} \left\{-i\,S_\sL\,{e_L(x)} \gamma_5
- {f_T(x)}\,\epsilon_\sT^{\rho\sigma}\gamma_\rho S_{\sT\sigma}
+ i\,{h(x)}\frac{[\slsh n_+,\slsh n_-]}{2} \right\}. 
\eeaq
The functions $e$, $g_T$ and $h_L$ are T-even, the
functions $e_L$, $f_T$ and $h$ are T-odd. They vanish for the matrix element
$\Phi$ in Eq.~\ref{PhiDIS} , but we keep them because in
the analogous situation of fragmentation functions, to be discussed
further down, they become very important.
The leading order evolution of $e$, $g_T$ and $h_L$
is known \cite{Evol} and for the non-singlet
case this also provides the evolution of the T-odd functions $e_L$,
$f_T$ and $h$ respectively, for which the operators involved differ only from
those of the T-even functions by a $\gamma_5$ matrix. 

The twist assignment is more evident by connecting these twist-3
functions to 
matrix elements of the form $\langle P,S \vert \overline
\psi_j(0)\,{\cal U}(0,\eta) \,iD_\sT^\alpha (\eta)\,{\cal
U}(\eta,\xi)\,\psi_i(\xi) \vert P,S \rangle$ via
the QCD equations of motion.

Of the twist-3 functions, $e$ and $h_L$ are chiral-odd,
while $g_T$ is chiral-even. Only these latter functions $g_T^q$,
reinstating flavor indices, appears in the polarized
DIS cross section in the structure function
$g_T(\xbj,Q^2)$ = $(g_1 + g_2)(\xbj,Q^2)$ = 
$\frac{1}{2}\sum_q e_q^2\,g_T^q(\xbj)$.
The calculation actually requires taking into account not only the
handbag diagram, shown as the left part in Fig.~\ref{fig1}, but also
diagrams in which an additional gluon emerges from the soft part.
These additional contributions with matrix elements containing
both quark and gluon fields can be rewritten into the twist-3
functions by the use of QCD equations of motion.

\section{Semi-inclusive leptoproduction}

\begin{figure}
\begin{center}
\epsfig{file=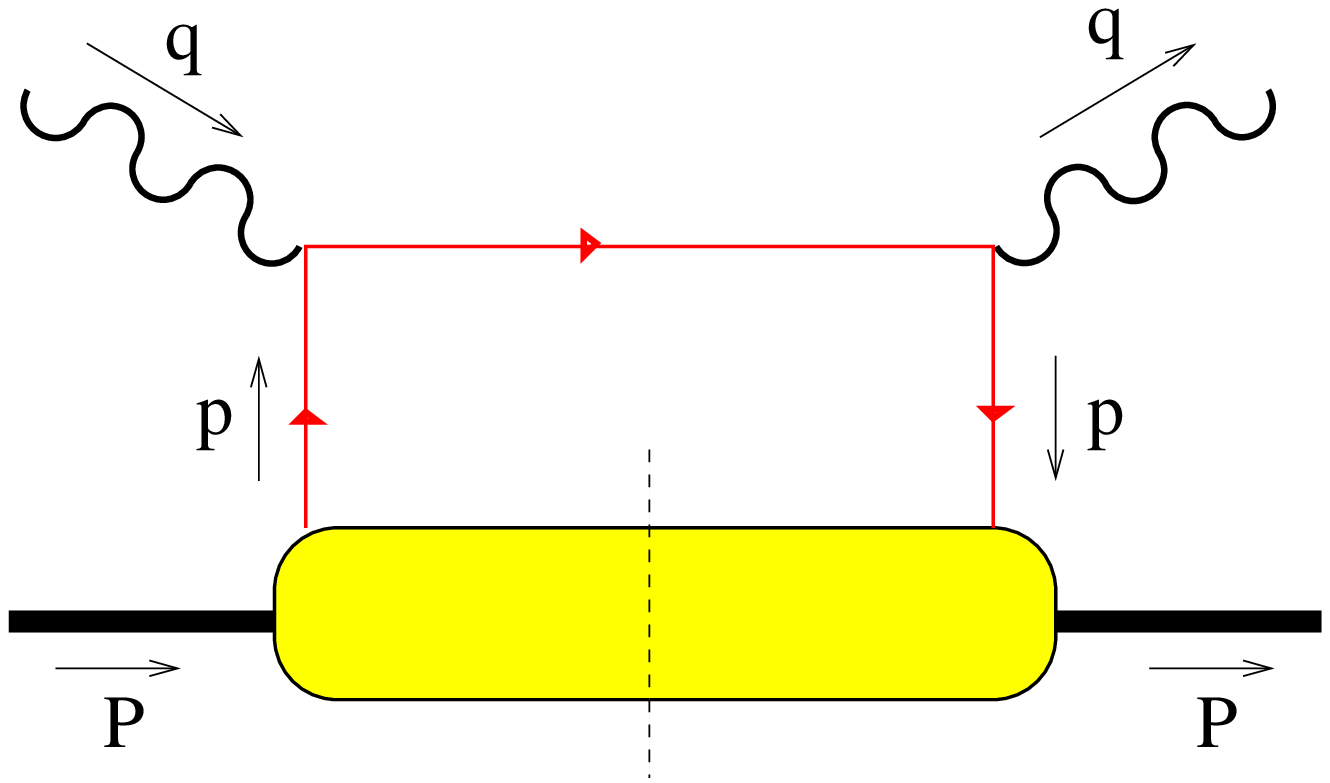,width=6cm} \hspace{2cm}
\epsfig{file=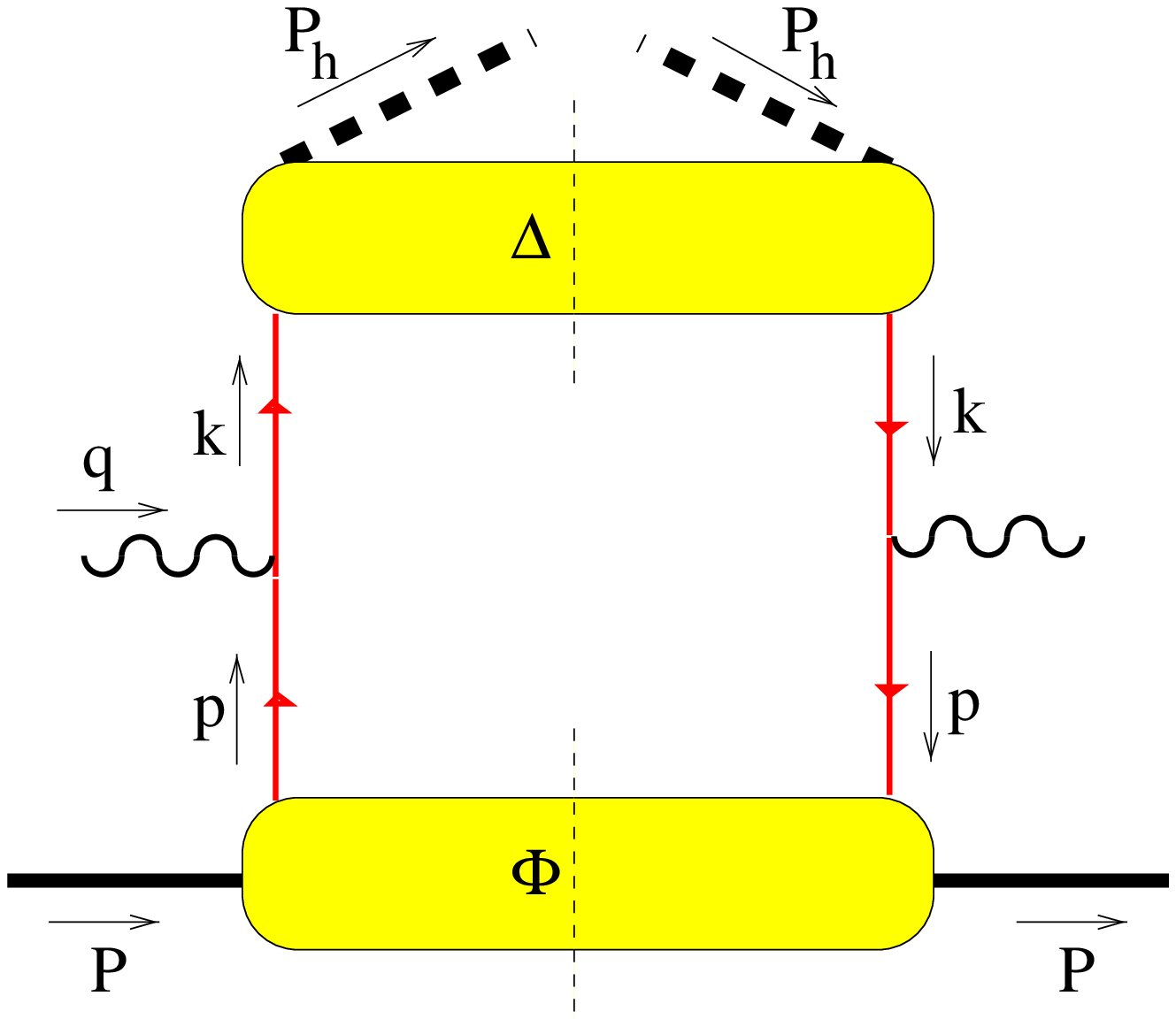, width=6cm}
\end{center}
\caption{The leading contributions to inclusive (left) and 1-particle 
inclusive (right) leptoproduction. \label{fig1}}
\end{figure}

Semiinclusive DIS (SIDIS), in particular one-particle inclusive DIS,
can be used for additional flavor
identification. Instead of weighing quark flavors with the quark charge
squared $e_q^2$ one obtains a weighting with $e_q^2\,D_1^{q\rightarrow
h}(z_h)$, where $D_1^{q\rightarrow h}$ is the usual unpolarized
fragmentation function
for a quark of flavor $q$ into hadron $h$, experimentally accessible at 
$z_h$ = $P\cdot P_h/P\cdot q$. In this contribution, we want to focus on 
the possibilities to study the intrinsic transverse momentum of partons, q
uarks and gluons. This is possible via azimuthal asymmetries, often appearing 
in combination with spin asymmetries. 

Azimuthal asymmetries appear in hard scattering processes with at least two
relevant hadrons. By relevant hadrons we mean hadrons
used as target or detected in the final state. 
Factorization crucially depends on the presence of a large energy scale in 
the process, such as the space-like momentum transfer squared $q^2 = -Q^2$ in 
leptoproduction. In this paper we will be concerned with functions that
appear in hard processes but which have, apart from such a hard scale, an
additional soft momentum scale, related to the transverse momentum 
of the partons. In one-hadron inclusive leptoproduction such a scale is
accessible 
because one deals with three momenta: the large momentum transfer $q$, the
target momentum $P$ and the momentum of the produced hadron $P_h$. The
noncollinearity at the quark level appears via $q_\sT$ = $q + \xbj\,P -
P_h/z_h$, where $\xbj =  Q^2/2P\cdot q$ and $z_h = P\cdot P_h/P\cdot q$ are
the usual semiinclusive scaling variables, at large $Q^2$ identified with
lightcone momentum fractions. The hadron momenta $P$ and $P_h$ define in
essence two lightlike directions\footnote{
For SIDIS the lightlike directions $n_\pm$ and lightcone
coordinates $a^\pm = a\cdot n_\mp$ are naturally
defined through hadron momenta $P$
and $P_h$, in which case the momentum transfer $q$ requires a transverse
component 
\begin{eqnarray*}
P = \frac{Q}{\xbj\sqrt{2}}\,n_+ + \frac{\xbj M^2}{Q\sqrt{2}}\,n_-,
\\
q = -\frac{Q}{\sqrt{2}}\,n_+ + \frac{Q}{\sqrt{2}}\,n_- + q_\st,
\\
P_h = \frac{M_h^2}{Z_h\,Q\sqrt{2}}\,n_+ + \frac{z_h\,Q}{\sqrt{2}}\,n_-.
\end{eqnarray*}
} 
$n_+$ and $n_-$, respectively. The soft 
scale is $Q_\sT^2 = -q_\sT^2$.

While the distribution functions in DIS in leading order
could be obtained from the lightcone
correlation function in Eq.~\ref{PhiDIS}, one encounters in SIDIS two types of
lightfront (where only one lightcone component vanishes) correlation functions
(see right part of Fig.~\ref{fig1}),
involving also transverse momenta of
partons as first pointed out by Ralston and Soper~\cite{RS79,TM95}
One part is relevant to treat quarks in a hadron
\be 
\Phi_{ij}(x,\bm p_T) =
\left. \int \frac{d\xi^-d^2\bm \xi_T}{(2\pi)^3}\ e^{ip\cdot \xi}
\,\langle P,S\vert \overline \psi_j(0) \psi_i(\xi)
\vert P,S\rangle \right|_{\xi^+ = 0},
\label{phi}
\ee
depending on $x=p^+/P^+$ and the quark transverse 
momentum $\bm p_\st$ in a target with $P_\st = 0$.
A second correlation function~\cite{CS82}
\be
\Delta_{ij}(z,\bm k_\st) =
\left. \sum_X \int \frac{d\xi^+d^2\bm \xi_\st}{(2\pi)^3} \,
e^{ik\cdot \xi} \langle 0 \vert \psi_i (\xi) \vert P_h,X\rangle
\langle P_h,X\vert\overline \psi_j(0) \vert 0 \rangle
\right|_{\xi^- = 0},
\label{delta}
\ee
describes fragmentation of a quark into a hadron. It depends
on $z = P_h^-/k^-$ and the quark transverse momentum $k_\st$ when one
produces a hadron with $P_{h\st} = 0$. A simple boost shows  that this is
equivalent to a quark producing a hadron with transverse  momentum $P_{h\perp}
= -z\,k_\st$ with respect to the quark. 

The parametrization of the $x$ and $p_\sT$ dependent correlators
becomes~\cite{RS79,MT-96,BM-98} 
\begin{eqnarray}
\Phi(x,\bm{p}_\sT) & = & 
\frac{1}{2}\,\Biggl\{
f_1(x ,\bm p^2_\sT)\,\slsh n_+ + 
f_{1T}^\perp(x ,\bm p^2_\sT)\, \frac{\epsilon_{\mu \nu \rho \sigma}\gamma^\mu 
n_+^\nu p_\sT^\rho S_{\sT}^\sigma}{M}
\nonumber \\ && \mbox{}  
- g_{1s}(x ,\bm p_\sT)\, \slsh n_+ \gamma_5
- h_{1T}(x ,\bm p^2_\sT)\,i\sigma_{\mu\nu}\gamma_5 S_{\sT}^\mu n_+^\nu
\nonumber \\ && \mbox{}
- h_{1s}^\perp(x ,\bm p_\sT)\,\frac{i\sigma_{\mu\nu}\gamma_5 p_\sT^\mu
n_+^\nu}{M} + h_1^\perp (x,\bm p^2_\sT) \, \frac{\sigma_{\mu\nu} p_\sT^\mu
n_+^\nu}{M}\Biggl\}.
\label{paramPhixkt}
\end{eqnarray}
We used the shorthand notation
$g_{1s}(x, \bm p_\sT) \equiv
S_\sL\,g_{1L}(x ,\bm p_\sT^2)
+ \frac{(\bpt\cdot\bm{S}_{\sT})}{M}\,g_{1T}(x ,\bm p_\sT^2)$,
and similarly for $h_{1s}^\perp$.
The parameterization contains two T-odd functions, the
functions $f_{1T}^\perp$ and $h_1^\perp$. The whole
treatment of the fragmentation functions is analogous with dependence on the
quark momentum fraction $z = P_h^-/k^-$ and $k_\sT$. In the notation for
fragmentation functions we 
employ the replacements $f_{\ldots} \rightarrow D_{\ldots}$, $g_{\ldots} 
\rightarrow G_{\ldots}$ and $h_{\ldots} \rightarrow
H_{\ldots}$. One also encounters a hat notation.

Again we give the matrix representation which is useful for the 
interpretation of the functions and which can be used to derive positivity
constraints~\cite{BBHM}. Including transverse momentum dependence one finds 
the production matrix $M^{\rm (prod)}$ being
\be
\left\lgroup \begin{array}{cccc}
f_1 + g_{1} &
\frac{\vert p_\st\vert}{M}\,e^{i\phi}\,g_{1T}&
\frac{\vert p_\st\vert}{M}\,e^{-i\phi}\,h_{1L}^\perp&
2\,h_{1} \\
& & & \\
\frac{\vert p_\st\vert}{M}\,e^{i\phi}\,g_{1T}^\ast&
f_1 - g_{1} &
\frac{\vert p_\st\vert^2}{M^2}e^{-2i\phi}\,h_{1T}^\perp &
-\frac{\vert
p_\st\vert}{M}\,e^{-i\phi}\,h_{1L}^{\perp\ast}\\ & & &
\\ \frac{\vert p_\st\vert}{M}\,e^{i\phi}\,h_{1L}^{\perp\ast}&
\frac{\vert p_\st\vert^2}{M^2}e^{2i\phi}\,h_{1T}^\perp &
f_1 - g_{1} &
-\frac{\vert p_\st\vert}{M}\,e^{i\phi}\,g_{1T}^\ast \\
& & & \\
 2\,h_{1}&
-\frac{\vert p_\st\vert}{M}\,e^{i\phi}\,h_{1L}^\perp&
-\frac{\vert p_\st\vert}{M}\,e^{-i\phi}\,g_{1T} &
f_1 + g_{1}
\end{array}\right\rgroup ,
\ee
to be compared with Eq.~\ref{prod1}. We have omitted here the T-odd functions
$f_{1T}^\perp$ and $h_1^\perp$ appearing as imaginary parts of $g_{1T}^{\mbox{}}$
and $h_{1L}^\perp$, respectively. Using time-reversal invariance all the
distribution functions appearing in this matrix are expected to be real,
leaving aside mechanisms such as discussed in Refs~\cite{Sivers90}. 
For $\Delta(z, k_\st)$, a correlator that contains explicit out-states
$\vert P_h,X\rangle$, T-reversal cannot be used, in contrast to $\Phi$,
a correlator that only contains
plane-wave hadron states~\cite{RKR71,HHK83,JJ93}.
Thus in that case one finds the two T-odd fragmentation 
functions~\cite{Collins93,MT-96} appearing
as imaginary parts of the complex off-diagonal ($p_\st$-dependent) functions
in the matrix representation. The matrix corresponds for $\Delta(z,k_\st)$ 
actually to a semi-definite quark decay matrix.

\section{Azimuthal asymmetries}

The possibility to access the full (transverse momentum dependent) spin
structure of the nucleon is in my opinion one of the most exciting
possibilities offered by 1-particle inclusive leptoproduction. It 
requires measurements of azimuthal asymmetries, which for the leading
cross sections often appear only in combination with spin asymmetries. 
The T-odd  fragmentation functions play a special role since they appear in
single-spin asymmetries.
Recent experimental results have clearly shown the presence of azimuthal
asymmetries including single-spin asymmetries~\cite{LEP,HERMES,SMC,HERMES2}

At measured $q_\sT$ one deals with a
convolution of two transverse momentum dependent functions, where the
transverse momenta of the partons from different hadrons combine to
$q_\sT$~\cite{RS79,MT-96,Boer-00}. A decoupling is achieved by studying cross
sections weighted with the momentum $q_\sT^\alpha$, leaving only the
directional (azimuthal) dependence. For the quark distribution and 
fragmentation functions this implies a weighing with the transverse
momentum, which is also important for proper theoretical handling of
the functions~\cite{Boer-00}. 

We give a few explicit examples. For that we
define weighted cross sections such as
\be
\int d^2\bm q_{T}\,\frac{Q_\st^2}{MM_h} \,\sin(2\phi_h^\ell)
\,\frac{d\sigma_{{OL}}}{d\xbj\,dy\,dz_h\,d^2\bm q_{T}}
\equiv
\left< \frac{Q_{T}^2}{MM_h} \,\sin(2\phi_h^\ell)\right>_{OL} ,
\ee
where the angle $\phi_h^\ell$ is the azimuthal angle between the hadron
production plane and the lepton scattering plane. The subscripts $O$, $L$
and $T$ are used to refer to polarization of lepton and target respectively.
The subsequently used angle $\phi_S^\ell$ is the azimuthal angle between
the transverse target spin vector and the lepton scattering plane.

For polarized targets, several azimuthal asymmetries arise already
at leading order. For example the following possibilities were
investigated in Refs~\cite{KM96,Collins93,Kotzinian95,TM95b}.
\bea
&&
\left< \frac{Q_\st}
{M} \,\cos(\phi_h^\ell-\phi_S^\ell)\right>_{LT} =
\nonumber \\ && \qquad
\frac{2\pi \alpha^2\,s}{Q^4}\,{\lambda_e\,\vert \bm S_\st \vert}
\,y(2-y)\sum_{a,\bar a} e_a^2
\,\xbj\,{g_{1T}^{(1)a}}(\xbj) {D^a_1}(z_h),
\label{asbas}
\\
&&
\left< \frac{Q_\st^2}{MM_h}
\,\sin(2\phi_h^\ell)\right>_{OL} =
\nonumber \\ && \qquad
-\frac{4\pi \alpha^2\,s}{Q^4}\,{\lambda}
\,(1-y)\sum_{a,\bar a} e_a^2
\,\xbj\,{h_{1L}^{\perp(1)a}}(\xbj) {H_1^{\perp(1)a}}(z_h),
\label{as2}
\\
&&
\left< \frac{Q_\st}{M_h}
\,\sin(\phi_h^\ell+\phi_S^\ell)\right>_{OT} =
\nonumber \\ && \qquad
\frac{4\pi \alpha^2\,s}{Q^4}\,{\vert \bm S_\st \vert}
\,(1-y)\sum_{a,\bar a} e_a^2
\,\xbj\,{h_1^a}(\xbj) {H_1^{\perp(1)a}}(z_h).
\label{finalstate}
\eea
The latter two are single spin asymmetries involving the T-odd fragmentation
function $H_1^{\perp (1)}$. The last one was the asymmetry proposed by
Collins~\cite{Collins93} as a way to access the transverse spin 
distribution function $h_1$ in pion production. In the weighted cross
sections, the socalled transverse moments show up, defined as
\be
f^{(n)}(x) = \int d^2p_\sT\,\left(\frac{\bm p_\sT^2}{2M^2}\right)^n
\,f(x,\bm p_\sT).
\ee
These transverse moments only depend on $x$ and are
contained in 
$\Phi_\partial^\alpha (x) \equiv
\int d^2 p_\sT\,\frac{p_\sT^\alpha}{M} \,\Phi(x,\bm p_\sT)$
which projects out the functions in $\Phi(x,\bm p_\sT)$ where $p_\sT$
appears linearly,
\beaq
\Phi_\partial^\alpha (x) & = &
\frac{1}{2}\,\Biggl\{
-g_{1T}^{(1)}(x)\,S_\sT^\alpha\,\slsh n_+\gamma_5
-S_\sL\,h_{1L}^{\perp (1)}(x)
\,\frac{[\gamma^\alpha,\slsh n_+]\gamma_5}{2}
\nonumber \\
&&\quad \mbox{}
-{f_{1T}^{\perp (1)}}(x)
\,\epsilon^{\alpha}_{\ \ \mu\nu\rho}\gamma^\mu n_-^\nu {S_\sT^\rho}
- i\,{h_1^{\perp (1)}}(x)
\,\frac{[\gamma^\alpha, \slsh n_+]}{2}\Biggr\} .
\label{Phid}
\eeaq

\section{The operator structure of transverse moments~\cite{HBM}}

To study the scale dependence we use the moments in both $p_\sT$ and $x$
of the $p_\st$-dependent functions, employ Lorentz  invariance and use the QCD
equations of motion. The moments in $x$ for leading (collinear)  distribution
functions (appearing for instance in inclusive leptoproduction) are related to
matrix  elements of twist two operators. On the other hand, for the transverse 
moments entering the azimuthal asymmetry expressions of interest, one finds 
relations to matrix elements of twist two {\em and\/} twist three operators, 
for which the evolution, however, is known. In the large $N_c$ limit this 
evolution becomes particularly simple.

We first invoke Lorentz invariance to
rewrite some functions discussed so far. All functions in $\Phi(x)$ and 
$\Phi_\partial^\alpha(x)$ involve nonlocal matrix elements of two quark 
fields. Before constraining
the matrix elements to the light-cone or lightfront only a limited number
of amplitudes can be written down. This leads to the following
Lorentz-invariance relations
\beaq
&&g_T  = g_1 + \frac{d}{dx}\,g_{1T}^{(1)}, \qquad
h_L = h_1 - \frac{d}{dx}\,h_{1L}^{\perp (1)},
\label{gTrel}
\\
&&f_T =  - \frac{d}{dx}\,f_{1T}^{\perp (1)}, \qquad
\ h =  - \frac{d}{dx}\,h_{1}^{\perp (1)}.
\label{rel4}
\eeaq
From these relations, it is clear that  
the $\bm p_\sT^2/2M^2$ moments of the $p_\sT$-dependent 
functions, appearing in $\Phi_\partial^\alpha(x)$, involve both twist-2
and twist-3 operators. 

Another useful set of functions is obtained as the difference between
the correlator $\Phi_D$, which via equations of motion is connected to
$\Phi^{\rm twist-3}$, and the correlator $\Phi_\partial$. 
This difference corresponds in
$A^+ = 0$ gauge to a correlator $\Phi_A$, involving $\langle P,S \vert
\overline \psi_j(0)\,{\cal U}(0,\eta) A_\sT^\alpha (\eta)\,{\cal
U}(\eta,\xi)\,\psi_i(\xi) \vert P,S \rangle$. Via the parametrisation of
$\Phi_A$ one defines 
interaction-dependent (tilde) functions,
\beaq
&&
x\,g_T(x) - \frac{m}{M}\,h_1(x)-g_{1T}^{(1)}(x) 
+ i\left[ x\,f_T(x) + f_{1T}^{\perp(1)}(x)\right]
\equiv x\,\tilde g_T(x)
+ ix\,\tilde f_T(x),
\nonumber \\ &&
\label{collinear1}
\\ &&
x\,h_L(x) - \frac{m}{M}\,g_1(x) 
+2\,h_{1L}^{\perp (1)}(x) - ix\,e_L(x)\equiv
x\,\tilde h_L(x) - ix\,\tilde e_L(x),
\nonumber \\ &&
\\ &&
x\,e(x) - \frac{m}{M}\,f_1(x) + i\left[
x\,h(x) +2\,h_1^{\perp (1)}(x)\right]\equiv
x\,\tilde e(x) + ix\,\tilde h(x).
\nonumber \\ &&
\label{collinear4}
\eeaq

Using the equations of
motion relations in Eqs.~(\ref{collinear1}) - (\ref{collinear4}) and the
relations based on Lorentz invariance in Eqs.~(\ref{gTrel}) - (\ref{rel4}),
it is straightforward to relate the various twist-3 functions and the
$\bm p_\sT^2/2M^2$ (transverse) moments of $p_\sT$-dependent
functions. The results for the functions are (omitting quark mass
terms)
\bea
&&
g_T(x) = 
\int_x^1 dy\ \frac{g_1(y)}{y}
+ \left[ \tilde g_T(x) - \int_x^1 dy\ \frac{\tilde g_T(y)}{y}\right],
\\ &&
\frac{g_{1T}^{(1)}(x)}{x} =
\int_x^1 dy\ \frac{g_1(y)}{y}
- \int_x^1 dy\ \frac{\tilde g_T(y)}{y},
\\[0.5cm] &&
h_L(x) = 
2x\int_x^1 dy\ \frac{h_1(y)}{y^2}
+ \left[ \tilde h_L(x) - 2x\int_x^1 dy\ \frac{\tilde h_L(y)}{y^2}\right],
\\ &&
\frac{h_{1L}^{\perp(1)}(x)}{x^2} =
-\int_x^1 dy\ \frac{h_1(y)}{y^2}
+ \int_x^1 dy\ \frac{\tilde h_L(y)}{y^2},
\\[0.5cm] &&
f_T(x) = 
\left[ \tilde f_T(x) - \int_x^1 dy\ \frac{\tilde f_T(y)}{y}\right] ,
\\ &&
\frac{f_{1T}^{\perp (1)}(x)}{x} =
\int_x^1 dy\ \frac{\tilde f_T(y)}{y},
\\[0.5cm] &&
h(x) = 
\left[ \tilde h(x) - 2x\int_x^1 dy\ \frac{\tilde h(y)}{y^2}\right],
\\ &&
\frac{h_{1}^{\perp(1)}(x)}{x^2} =
\int_x^1 dy\ \frac{\tilde h(y)}{y^2} .
\eea
The relation for $g_T$ is just the Wandzura-Wilczek result~\cite{WW}, having
a complete analogue in the relation for $h_L$ discussed in Ref.~\cite{JJ92}.
In slightly different form the result for $g_{1T}^{(1)}$ has been discussed
in Ref.~\cite{BKL}.
Actually, we need not consider the T-odd functions separately. They can be
simply considered as imaginary parts of other functions, when we allow complex
functions. In particular one can expand the correlation functions into
matrices in Dirac space~\cite{BBHM} to show that the relevant combinations are 
$(g_{1T} - i\,f_{1T}^{\perp})$ which we can treat together as one complex 
function $g_{1T}$. Similarly we can use complex functions
$(h_{1L}^\perp + i\,h_1^\perp)$ $\rightarrow$ $h_{1L}^\perp$,
$(g_T + i\,f_T)$ $\rightarrow$ $g_T$,
$(h_L + i\,h)$ $\rightarrow$ $h_L$,
$(e + i\,e_L)$ $\rightarrow$ $e$. The functions $f_1$, $g_1$ and $h_1$ remain
real, they don't have T-odd partners. 

\section{Evolution of transverse moments~\cite{HBM}}

As mentioned the evolution of the twist-2 functions and the tilde functions in
known. The twist-2 functions have an autonomous evolution of the form
\beq
\frac{d}{d\tau} \,f(x,\tau) = \frac{\alpha_s(\tau)}{2\pi}
\,\int_x^1 \frac{dy}{y} \ P^{[f]}\left(\frac{x}{y}\right)\,f(y,\tau),
\eeq
where $\tau$ = $\ln Q^2$ and $P^{[f]}$ are the splitting functions. 
In the large $N_c$ limit, also the tilde functions have an autonomous
evolution. Using the known splitting functions and the
relations given in the previous section, we then find the evolution of the
transverse moments,
\beaq
&&
\frac{d}{d\tau}\,g_{1T}^{(1)}(x,\tau) 
= \frac{\alpha_s(\tau)}{4\pi}\,N_c\int_x^1 dy\,\Biggl\{
\left[\frac{1}{2}\,\delta(y-x) + \frac{x^2+xy}{y^2(y-x)_+}\right]
\,g_{1T}^{(1)}(y,\tau)
\nonumber \\ && \hspace{8cm}
+ \frac{x^2}{y^2}\,g_1(y,\tau)\Biggr\} ,
\\ && 
\frac{d}{d\tau}\,h_{1L}^{\perp (1)}(x,\tau) 
= \frac{\alpha_s(\tau)}{4\pi}\,N_c\int_x^1 dy\,\Biggl\{
\left[\frac{1}{2}\,\delta(y-x) + \frac{3x^2-xy}{y^2(y-x)_+}\right]
\,h_{1L}^{\perp (1)}(y,\tau)
\nonumber \\ && \hspace{8cm}
-\frac{x}{y}\,h_1(y,\tau)\Biggr\}.
\eeaq
One can also analyse the fragmentation functions or use some specific
reciprocity relations~\cite{HBM}.
Furthermore, we note that apart from a
$\gamma_5$ matrix the operator structures of the T-odd functions 
$f_{1T}^{\perp (1)}$ and $h_1^{\perp (1)}$ are in fact the same 
as those of $g_{1T}^{(1)}$ and $h_{1L}^{\perp (1)}$ (or as mentioned before,
they can be considered as the imaginary part of these functions~\cite{BBHM}).
With these ingredients one immediately obtains for the non-singlet functions
the (autonomous) evolution of the T-odd fragmentation functions. 
In particular we obtain for
the Collins fragmentation function (at large $N_c$),
\bea
&&\frac{d}{d\tau} \,z H_{1}^{\perp (1)}(z,\tau) 
= \frac{\alpha_s}{4\pi}\; N_c\; \int_z^1 dy\, \left[
\frac{1}{2}\,\delta(y-z) + \frac{3y-z}{y(y-z)_+} \right] 
\,y H_{1}^{\perp (1)}(y,\tau),
\nonumber \\ &&
\eea
which should prove useful for the comparison of data on Collins
function asymmetries from  different experiments, performed at different
energies.

\section{Summary}

In this overview we have presented the $p_\st$-dependent quark distribution
and fragmentation functions that appear in azimuthal asymmetries in
one-particle inclusive leptoproduction. We have shown how appropriately
weighted cross sections contain socalled transverse moments, including
the T-odd fragmentation functions relevant in single-spin asymmetries. Using
Lorentz invariance and the QCD equations of motion, these transverse
moments can be related to the twist-2 
and twist-3 functions appearing in inclusive leptoproduction.
This also enables us to find evolution equations for the $p_\sT$-dependent
functions that appear in azimuthal asymmetries

\section*{Acknowledgments}

The work reported here is part of the research program of the Dutch
Foundation for Fundamental Research on Matter (FOM) and is partially
funded by the European Commission IHP program under contract 
HPRN-CT-2000-00130


\section*{References}

\begin{thebibliography}{99}

\bibitem{BBHM}
A. Bacchetta, M. Boglione, A. Henneman and P.J. Mulders,
Phys.~Rev.~Lett.~85 (2000) 712.

\bibitem{JJ92}
R.L. Jaffe and X. Ji, Nucl. Phys.~B 375 (1992) 527.

\bibitem{Soffer95}
J. Soffer, Phys. Rev. Lett.~74 (1995) 1292.

\bibitem{Evol} 
see e.g.
A.V. Belitsky, Lectures given at the XXXI PNPI
Winter School on Nuclear and Particle Physics, St. Petersburg, Repino,
February, 1997, hep-ph/9703432.

\bibitem{RS79}
J.P. Ralston and D.E. Soper, Nucl. Phys.~B 152 (1979) 109.

\bibitem{TM95}
R. D. Tangerman and P.J. Mulders, Phys. Rev.~D 51 (1995) 3357

\bibitem{CS82}
J.C. Collins and D.E. Soper, Nucl. Phys.~B 194 (1982) 445.

\bibitem{MT-96}
P.J. Mulders and R.D. Tangerman, Nucl.~Phys.~B 461 (1996) 197 and
Nucl.~Phys.~B 484 (1997) 538 (E).

\bibitem{BM-98}
D. Boer and P.J. Mulders, Phys.~Rev.~D 57 (1998) 5780.

\bibitem{Sivers90}
Possible T-odd effects could arise from soft initial state interactions
as outlined in 
D. Sivers, Phys. Rev.~D 41 (1990) 83 and Phys. Rev.~D 43 (1991) 261
and M. Anselmino, M. Boglione and F. Murgia, Phys. Lett.~B 362 (1995) 164.
Also gluonic poles might lead to presence of T-odd functions, see
N. Hammon, O. Teryaev and A. Sch\"afer, Phys.~Lett.~B 390 (1997) 409
and D. Boer, P.J. Mulders and O.V. Teryaev, Phys.~Rev.~D 57 (1998) 3057.

\bibitem{RKR71}
A. De R\'ujula, J.M. Kaplan and E. de Rafael, 
Nucl. Phys.~B 35 (1971) 365.

\bibitem{HHK83}
K. Hagiwara, K. Hikasa and N. Kai, Phys. Rev.~D 27 (1983) 84.

\bibitem{JJ93}
R.L. Jaffe and X. Ji, Phys. Rev. Lett.~71 (1993) 2547.

\bibitem{Collins93}
J. Collins, Nucl. Phys.~B 396 (1993) 161.

\bibitem{LEP}
E. Efremov, O.G. Smirnova and L.G. Tkatchev,
Nucl. Phys. Proc. Suppl.~B 79 (1999) 554.

\bibitem{HERMES}
A. Airapetian {\it et al.}, HERMES Collaboration, Phys.~Rev.~Lett.~84
(2000) 4047.

\bibitem{SMC}
A. Bravar, Nucl. Phys. Proc. Suppl.~B 79 (1999) 521.

\bibitem{HERMES2}
H. Avakian, Nucl. Phys. Proc. Suppl.~B 79 (1999) 523.

\bibitem{Boer-00} 
D. Boer, Phys.~Rev.~D 62 (2000) 094029 and hep-ph/0102071.

\bibitem{KM96}
A.M. Kotzinian and P.J. Mulders, Phys. Rev.~D 54 (1996) 1229; 
A.M. Kotzinian and P.J. Mulders, Phys. Lett.~B 406 (1997) 373.

\bibitem{Kotzinian95}
A. Kotzinian, Nucl. Phys. {\bf B 441} (1995) 234.

\bibitem{TM95b}
R.D. Tangerman and P.J. Mulders, Phys. Lett. {\bf B352} (1995) 129.

\bibitem{HBM}
A. Henneman, D. Boer and P.J. Mulders, hep-ph/0104271.

\bibitem{WW}
S. Wandzura and F. Wilczek, Phys. Rev.~D 16 (1977) 707.

\bibitem{BKL}
A.P. Bukhvostov, E.A. Kuraev and L.N. Lipatov, Sov. Phys. JETP~60
(1984) 22.

\end{thebibliography}
\end{document}